\documentclass[12pt]{article}
\usepackage{graphics,cite,amssymb,epsfig,float,psfrag}
\usepackage[usenames,dvips]{color}
\usepackage{rotating}

\begin{document}

\newcommand{\lsim}{\raisebox{-0.13cm}{~\shortstack{$<$ \\[-0.07cm] $\sim$}}~}
\newcommand{\gsim}{\raisebox{-0.13cm}{~\shortstack{$>$ \\[-0.07cm] $\sim$}}~}
\newcommand{\ra}{\rightarrow}
\newcommand{\lra}{\longrightarrow}
\newcommand{\ee}{e^+e^-}
\newcommand{\gam}{\gamma \gamma}
\newcommand{\nn}{\noindent}
\newcommand{\non}{\nonumber}
\newcommand{\beq}{\begin{eqnarray}}
\newcommand{\eeq}{\end{eqnarray}}
\newcommand{\s}{\smallskip}
\def\NPB{Nucl. Phys. B}
\def\PLB{Phys. Lett. B}
\def\PRL{Phys. Rev. Lett.}
\def\PRD{Phys. Rev. D}
\def\ZPC{Z. Phys. C}
\baselineskip=16pt

\rightline{LPT--Orsay 06/060}
\rightline{LAL--Orsay 06/144}
\rightline{hep-ph/0610173}
\rightline{October 2006}

\vspace{0.7cm}

\begin{center}

{\Large {\bf Resolving the $\mathbf{A_{FB}^b}$ puzzle in an extra dimensional 
}}

\vspace{0.2cm}

{\Large {\bf model with an extended gauge structure}}

\vspace{0.7cm}

{\large Abdelhak Djouadi$^1$, Gr\'egory Moreau$^1$, Fran\c cois Richard$^2$}

\vspace{0.7cm}

$^1$ Laboratoire de Physique Th\'eorique, CNRS and Universit\'e Paris--Sud, \\
B\^at. 210, F--91405 Orsay Cedex, France.


\vspace{0.2cm}

$^2$ Laboratoire de l'Acc\'el\'erateur Lin\'eaire, IN2P2--CNRS et Universit\'e 
de Paris--Sud, \\ B\^at. 200, BP 34, F--91898 Orsay Cedex, France.

\end{center}

\vspace{0.5cm}

\begin{abstract} 

It is notorious that, contrary to all other precision electroweak data, the
forward--backward asymmetry for $b$ quarks $A_{FB}^b$ measured in $Z$  decays at
LEP1 is nearly three standard deviations away from the predicted value in the
Standard Model; significant deviations also occur in measurements of the
asymmetry off the $Z$ pole. We show that these discrepancies can be resolved in
a variant of the Randall--Sundrum extra--dimensional model in which the gauge
structure is extended  to ${\rm SU(2)_L\! \times\! SU(2)_R\! \times\! U(1)_{X}}$
to allow for relatively light Kaluza--Klein excitations of the gauge bosons.  In
this scenario, the  fermions are localized differently along the extra 
dimension, in order to generate the fermion mass hierarchies, so that the  
electroweak interactions for the heavy third generation fermions are naturally
different from the light fermion ones. We show that the mixing between the $Z$
boson with the Kaluza--Klein excitations  allows to explain the $A_{FB}^b$
anomaly without affecting (and even improving) the agreement of the other
precision observables, including the $Z \to b\bar b$ partial decay width,  with
experimental data. Some implications of this scenario for the ILC are 
summarized.  

\end{abstract}

\newpage

\subsection*{1. Introduction} 

The Standard Model (SM) of the strong and electroweak interactions of elementary
particles \cite{GSW} has brilliantly passed almost all the experimental  tests
to date. These tests, performed at the per--mille level accuracy,  have probed
the quantum corrections and the structure of the ${\rm SU(3)_C \times SU(2)_L
\times U(1)_Y}$ local symmetry \cite{PDG,LEPex}.  However, it is notorious that
there is an observable for which the measured value significantly differs from
the one predicted in the context of the SM.  The forward--backward (FB)
asymmetry for $b$ quark jets $A_{FB}^b$ \cite{AFB-SM} measured in $Z$ boson
decays at LEP1 provides, together with the longitudinal asymmetry $A_{LR}$
measured at the SLC, the most precise individual measurement of the electroweak
mixing angle $\sin^2\theta^{\rm lep}_{\rm eff}$, but the result is 2.8 standard
deviations away from the predicted value and the two individual measurements
differ by more than three standard deviations. In fact, the $A_{FB}^b$ anomaly
is present not only in the data collected at the $Z$ boson pole, but also a few
GeV above and, to a lesser extent, a few GeV below the $Z$ pole. [There are also
measurements of $A_{FB}^b$ at energies below (from PEP to TRISTAN)
\cite{AFB-below,Klaus,AFB-mix} and far  above (LEP2) \cite{AFB-above} the 
$Z$--pole which are in a reasonable  agreement with the SM expectations.]\s 

This situation has led to speculations about a possible signal of New Physics
beyond the SM in the $Zb\bar b$ vertex\footnote{Of course, the possibility that
this discrepancy could be simply   due to some unknown experimental problem or
inaccuracy or to a large statistical fluctuation cannot be excluded
\cite{AFB-exp}.}.  However, it turns out that this discrepancy cannot be easily
explained without affecting the $Z\to b\bar{b}$ partial decay width, which has
also been very precisely measured at LEP1 and found to be compatible with the
SM expectation, and the FB longitudinal asymmetry $A_{FB,LR}^b$
measured at the SLC, although the corresponding experimental error is larger
\cite{LEPex}.  The basic reason for this difficulty lies in the fact that a
very large correction, ${\cal O}(30\%)$, is needed to alter the initially small
right--handed component of the $b$--quark coupling to $Z$ bosons without
affecting significantly the left--handed $b$--quark coupling as well as the
couplings of all other light fermions.  Note that this large effect can
possibly not be generated through radiative corrections in reasonable models
[this is, for instance, the case in supersymmetric extensions of the SM where
loop contributions of relatively light partners of the top quark and the $W$
boson \cite{AFB-SUSY} cannot account for the discrepancy] and one has to resort
to large tree--level effects to explain the anomaly.\s

Several New Physics models fulfilling theses conditions have been proposed in
the literature, though. Three examples of such models are variations of
left--right symmetric models in which a new gauge boson $Z'$ occurs and has
interactions only with third generation fermions \cite{AFB-Z'}, models
involving additional exotic (mirror) bottom--like quarks which strongly mix
with the $b$--quarks \cite{AFB-mix,AFB-mix-old} and models where the electroweak
symmetry breaking is induced by a new strongly interacting sector coupled to
the SM fields \cite{AFB-strong}. In most of the models listed above, a kind of
non--universality in which the third generation fermions are treated
differently from the ones of the first and second generations is assumed. In
most cases, this assumption is made in an ad--hoc way and specifically to cure
the $A_{FB}^b$ anomaly.\s 

Recently, it has been shown \cite{AFB-Rold} that in different scenarios beyond
the SM, such as Technicolor scenarios, Little Higgs theories, Higgsless models
and models where the composite Higgs boson arises as a pseudo--Goldstone boson
[which are related to theories in five--dimensional Anti-de-Sitter space via
the AdS/CFT correspondence], one can invoke a custodial O(3) symmetry which
protects at the same time the $\rho$ parameter and the $Zb_L\bar b_L$ coupling.
Such an approach might also account for the required shift of the $Z b_R\bar
b_R$ coupling to explain the $A_{FB}^b$ anomaly, if one allows for a deviation
of the $Z \to b \bar b$ partial decay width; this point will be discussed
later. The implications of this O(3) symmetry have been subsequently discussed
within Higgsless models \cite{Csaki} and gauge-Higgs unification scenarios
\cite{Carena}.\s

In this paper, we point out that the discrepancy between the measured value of
$A_{FB}^b$ and the theoretical prediction can be resolved in the context of
variants of the Randall--Sundrum (RS) extra--dimensional model \cite{RS} in
which the fermion and bosonic fields are propagating in the bulk. These models,
besides the fact that they explain the large hierarchy between the Planck and
TeV scales without introducing any fundamentally new energy scale in the
theory, allow for the unification of the gauge coupling constants at a high
energy Grand Unification scale \cite{UNI-RS}, provide a solution for proton
decay and have a viable candidate of Kaluza--Klein type for the dark matter of
the universe \cite{LZP2}. They also have the additional attractive feature of
providing a geometrical explanation of the large mass hierarchies prevailing
among SM fermions \cite{RSloc,HSquark,GNeubertA}.  Indeed, when the SM fermions
are localized differently along the extra dimension depending on their nature,
their different wave functions which overlap with the Higgs boson [which is
still localized on the TeV--brane for its mass to be protected] generate large
hierarchical patterns among their effective four-dimensional Yukawa couplings. 
One can then naturally obtain electroweak interactions for the heavy third
generation fermions that are different from the ones of the light fermions.\s 

More specifically, we will work in a variant of the RS model proposed in 
Ref.~\cite{ADMS} where the electroweak symmetry is enhanced to the left--right 
gauge symmetry ${\rm SU(2)_L\! \times\! SU(2)_R\! \times\! U(1)_{X}}$ in which 
the bulk fields are embedded, the right--handed fermions
being promoted to ${\rm SU(2)_R}$ isodoublets. With this symmetry, the 
high--precision electroweak measurements can be nicely fitted while keeping the
masses of the first Kaluza--Klein (KK) excitation modes of the gauge bosons
rather low, $M_{KK}$ of order of a few TeV, which is required if the gauge
hierarchy problem is to be addressed within the RS model with bulk matter. The
ordinary $Z$ boson will then mix with its KK excitations and those of the new
$Z'$ boson, with the possibility of choosing their effective couplings in such 
a way that {\it only} the overall $Z$ couplings to third generation fermions are
significantly altered. This would provide an explanation for the
$A_{FB}^b$ anomaly, while keeping the electroweak observables involving light
fermions as well as the $Z \to b \bar b$ decay width, unaltered for $M_{KK} =
{\cal O}$\,(TeV). In this context, we propose two scenarios: one in which the 
U(1) group is ${\rm U(1)_{\rm B-L} }$ (as originally discussed in 
Ref.~\cite{ADMS}) and a second in which the right--handed $b$--quark has 
isospin $I_{3R}^{b_R}\!=\!+\!\frac12$ under the ${\rm  SU(2)_R}$ group. While 
the $A_{FB}^b$ anomaly on the $Z$ pole is resolved in both scenarios, the 
experimental data for the asymmetry off the $Z$ pole are reproduced only in 
the second scenario.

The paper is organized as follows. In the next section, we briefly describe the
extra--dimensional model of Ref.~\cite{ADMS} based on the ${\rm SU(2)_L \times
SU(2)_R \times U(1)_{X}}$ symmetry and summarize the main features which allow
for relatively light KK states; the role of the fermions and their couplings to
the $Z$ boson are then summarized.  In section 3, we discuss two scenarios in
which the choice of $Z$--fermion couplings allows to explain the $A_{FB}^b$
anomaly at energies on the $Z$ pole and reproduce the value of the $Z \to b\bar
b$ partial decay width, without affecting the other precision measurements.  In
section 4, we analyze the asymmetry off the $Z$ pole and show that the agreement
between theory and experiment is satisfactory only in the scenario where the
right--handed $b$--quark has isospin $I_{3R}^{b_R} =+\frac12$ under the ${\rm
SU(2)_R}$ group. In section 5, we discuss the implications of this  scenario at
the ILC.  A brief conclusion is presented in section 6.

\subsection*{2. The physical set--up} 
\label{sub:context}

\subsubsection*{2.1 Theoretical framework} 
\label{sub:theory}

We consider the higher--dimensional Randall--Sundrum scenario in which the
Standard Model fields are propagating in the bulk like gravity, except for the
Higgs boson which remains confined on the TeV--brane in order not to
re--introduce a gauge hierarchy. On the Planck--brane, the gravity scale is
equal to the reduced Planck mass $M_{P}= 2.44\times 10^{18}$ GeV, whereas the
effective gravity scale on the TeV--brane, $M_{\star}=wM_{P}$, is suppressed by
the exponential ``warp" factor $w={\rm exp} (-\pi kR_{c})$ where $1/k$ is the
curvature radius of the anti--de Sitter space and $R_c$ the compactification
radius. For a small extra dimension, $R_{c} \simeq 11/k$ with $k$ being close
to $M_{P}$, one finds $w\sim 10^{-15}$ so that $M_{\star}\!=\!{\cal O}(1)$ TeV,
thus addressing the gauge hierarchy problem.  For this $R_c$ value, the
five--dimensional gravity scale $M_5$ is close to the effective
four--dimensional gravity scale $M_{P}$.\s

The mass of the first Kaluza--Klein excitations of the SM gauge bosons, $M_{KK}
=M_{\gamma^{(1)}}=M_{g^{(1)}} \approx M_{Z^{(1)}} \approx M_{W^{\pm(1)}}$,
reads $M_{KK}=2.45 kw=2.45 kM_{\star}/M_{P} \sim M_{\star} = {\cal O}$(TeV)
and, since this parameter is more ``physical", we adopt it as our free input 
parameter instead of $k$. The maximal value of $M_{KK}$ is fixed by $kR_{c}$ and
the theoretical consistency bound on the five--dimensional curvature scalar by
$|R_5|=|-20 k^2|<M_5^2$ leading to $k<0.105 M_{P}$; one can thus take a maximal
value of $M_{KK} \sim 10$ TeV which corresponds to the value $kR_{c}=10.11$. 
In fact, there exist a severe indirect bound of typically $M_{KK} \gtrsim 10$
TeV originating from electroweak precision data \cite{Burdman}. In order to
soften this bound down to a few TeV and, thus, to allow for a solution of the
gauge hierarchy problem, several scenarios have been proposed.  For instance,
scenarios with brane--localized kinetic terms for fermions  and
gauge bosons \cite{BraneFB} allow to lower the previous bound down to a few TeV
\cite{EWB}. Another possibility, that we retain here, is the model of
Agashe et al. \cite{ADMS} in which the electroweak gauge symmetry is enhanced
to a left--right ${\rm SU(2)_L \times SU(2)_R \times U(1)_{X}}$ structure in
the bulk and where right--handed fermions are promoted to ${\rm SU(2)_R}$
doublet fields, the new doublet components having no zero mode. The usual SM
electroweak symmetry is recovered after the breaking of both ${\rm SU(2)_{R}}$
and ${\rm U(1)_{X}}$ on the Planck--brane, with possibly a small breaking of
${\rm SU(2)_R}$ in the bulk.\s

As mentioned in the introduction, the RS model with bulk matter provides
naturally a geometrical interpretation of quark/lepton mass hierarchies
\cite{RSloc}.  The idea is to localize the SM fermions differently along the
extra dimension, depending on their nature. Then, the overlapping of their
different wave functions with the Higgs boson generates large hierarchical
patterns among the effective four--dimensional Yukawa coupling constants.  In
this mass model, each five--dimensional fermion field $\Psi_{i}$ [$i=\{1,2,3\}$
is the family index in the interaction basis] couples to a distinct mass $m_{i}$
in the fundamental theory as $\int d^{4}x \int dy \sqrt{G} m_{i} \bar{\Psi}_{i}
\Psi _{i}$  where $G$ is the determinant of the RS metric and $y$ parameterizes
the fifth dimension. The localization of fermions along the extra dimension is
fixed by the dependence of mass $m_{i}$ on $y$. One can take $m_{i}= {\rm
sign}(y) c_{i} k$  \cite{Tamvakis} where $c_{i}$ are dimensionless parameters.  
Then the fields decompose as 
$\Psi _{i}(x^{\mu },y)= \sum_{n=0}^{\infty }\psi_{i}^{(n)}(x^{\mu })
f_{n}^{i}(y)$, with $n$ labeling the tower of KK excitations, admitting as a
solution for the zero mode wave function $f_{0}^{i}(y)=\exp [(2-c_{i})k|y| ]/
N_{0}^{i}$, where $N_{0}^{i}$ is a normalization factor. The Yukawa 
interactions with the Higgs boson $H$ read then 
\begin{equation} 
{\cal S}_{\rm Yukawa} =
\int d^5x \sqrt{G} \left( Y_{ij}^{(5)} H \bar \Psi_{+ i} 
\Psi_{- j} + {\rm h.c.} \right) = \int d^4x M_{ij} \bar \psi_{L i}^{(0)} 
\psi_{R j}^{(0)} + {\rm h.c.} + \dots  
\label{eq:Yuk} 
\end{equation} 
where $Y_{ij}^{(5)}$ are the five--dimensional Yukawa coupling constants 
and the dots stand for KK mass terms. The effective fermion mass matrix 
is obtained after the integration 
\begin{equation} 
M_{ij} = \int dy \ \sqrt{G} \ Y_{ij}^{(5)} \ H f_0^i(y) f_0^j(y).  
\label{eq:MassMatrix} 
\end{equation} 
The dimensionful $Y_{ij}^{(5)}$ couplings can be chosen almost universal so 
that the quark/lepton mass hierarchies are essentially generated through the  
overlap between $f_0^{i,j}(y)$ and $H$ along $y$. Remarkably, large fermion 
mass hierarchies can be created for fundamental parameters $m_i$ all of the 
order of the unique scale of the theory, $k \sim M_5 \sim M_{P}$.

\subsubsection*{2.2 The couplings to fermions} 
\label{sub:Zcoupl}

Let us now discuss the fermions couplings to the $Z$ boson in the context of 
this extended left--right symmetric model. For each fermion $f_{L/R}$ of 
left/right--handed chirality, the couplings to the $Z$ boson which in the SM 
read 
\begin{equation} 
Q_Z^{f_{L/R}} = I_{3L}^{f_{L/R}} - Q^f_{\gamma} \sin^2 \theta_W, 
\end{equation}
where $I_{3L}^{f_L} =\pm \frac12$ and $I_{3L}^{f_R}=0$ are the third components 
of weak isospin and $Q_\gamma^f$ the electric charge of the fermion $f$. One 
has to add the contributions due to the mixing between the $Z$ boson and its 
excitations and those of the new $Z'$ boson, which is a superposition of the 
state $\widetilde W^3$ associated to the ${\rm SU(2)_R}$ group and $\widetilde 
B$ associated to ${\rm U(1)_{X}}$; the other orthogonal state is the SM 
hypercharge boson $B$. In terms of the coupling constant $g_{Z'}$ of this new 
$Z'$ and the charges $Q_{Z'}^{f_{L/R}}$ of the fermion $f$ and the Higgs boson
$Q_{Z'}^H$, the additional contributions to the $Z$ boson coupling to left-- 
and right--handed fermions reads 
\begin{equation} 
\frac{\Delta Q_Z^{f_{L/R}}}{Q_Z^{f_{L/R}}} = \frac{M_Z^2}{(0.4 M_{KK})^2}
\left[ \left( 1+\frac{1}{4F(c_{f_{L/R}})} \left[ 1-\frac{1}{k \pi R_c} \right]
\right) + \frac{g^2_{Z'} Q_{Z'}^{f_{L/R}} Q_{Z'}^H}{g^2_Z Q_Z^{f_{L/R}}
Q_Z^H}  \right] F(c_{f_{L/R}}), 
\label{Z'charges} 
\end{equation}
where $g_{Z'}$ is the $Z^{\prime}$ coupling constant. The first term comes from
the mixing between the $Z$ boson with its KK excitations $Z^{(n)}$, whereas the
second term is due to the mixing with the $Z^{\prime(n)}$ excitations; the $Z'$
boson is coupled to a Planckian vev on the ultraviolet brane, which mimics a
$(-,+)$ boundary condition to a good approximation, so that it possesses no
zero mode. In eq.~(\ref{Z'charges}), the $Z'$ charges are given, in terms of
the new mixing angle $\theta'$, the isospin number $I_{3R}^{f_{L/R}}$ under the 
${\rm SU(2)_R}$ gauge group, and the SM hypercharge $Y^{f_{L/R}}$ by 
\begin{equation} 
Q_{Z'}^{f_{L/R}} = I_{3R}^{f_{L/R}} - Y^{f_{L/R}} \sin^2 \theta' 
\end{equation} 
where the relation between the hypercharge and the charge $Q_X$ under the 
${\rm U(1)_X}$ group is 
\begin{equation} 
Y^{f_{L/R}}= Q_X^{f_{L/R}} + I_{3R}^{f_{L/R}} = Q_\gamma^f - I_{3L}^{f_{L/R}}, 
\label{eq:Ycond} 
\end{equation} 
In the previous equation, $\sin \theta' \equiv \tilde g' /g_{Z'}$ and 
$g_{ Z^\prime}^2 = \tilde g^2 + \tilde g^{\prime 2}$, where $\tilde g$ and 
$\tilde g'$ are, respectively, the ${\rm SU(2)_ R}$ and ${\rm U(1)_{X}}$ 
couplings; the coupling $g'$ of the SM ${\rm U(1)_Y}$ group reads $g'=\tilde g 
\tilde g' /g_{Z'}$. From these relations, one easily derive the following 
relations: 
\begin{eqnarray} 
2 \tilde g^{\prime 2}/ g_{Z'}^2 = 1 \pm \sqrt{1- (2 g'/g_{Z'}) ^2},
\ \ \ 
2 \tilde g^2/ g_{Z'}^2=1 \mp \sqrt{1-( 2 g'/ g_{Z'}) ^2} .
\end{eqnarray} 

Hence, if the charge of the U(1) group is $Q_X=\frac12 (B-L)$ with $B$ and $L$ 
being respectively the baryon and lepton numbers, that is exactly as in 
Ref.~\cite{ADMS}, one has $I_{3L}=-Y$ and $I_{3R}=Y$ for the charges 
of the neutral Higgs boson so that $Q_{Z'}^H / Q_Z^H \simeq -1$ in the
approximation that we will consider later: $\sin^2 \theta' \ll 1$.\s

Finally, the function  $F(c)$  in eq.~(\ref{Z'charges}) reads
\begin{equation}
F(c)= \frac{1}{1-e^{k \pi R_c (2 c -1)}} \frac{1-2c}{3-2c} 
\bigg[ \frac{5-2c}{4(3-2c)} - \frac{k \pi R_c}{2}  \bigg].   
\label{gen}
\end{equation}
Given the exponential term in eq.~(\ref{gen}), $F(c)$ goes rapidly to zero above
the value $c=0.5$; it remains everywhere negative and has no singularities and,
in particular, it is finite at $c=0.5$ with $F(0.5) \simeq - 0.25$. The
dependence of the $F$ function on the fermion $c$ parameters is due  to the fact
that the effective four--dimensional couplings between KK gauge  bosons and zero
mode fermions depend on the fermion localization which is given by the $c$
parameters.\s

The couplings of the first KK excitations of the $Z$ and $Z'$ bosons to SM
fermions have expressions that can be found in Ref.~\cite{RizzoUniv} for
instance. The ratio $Q(c)$ ($Q'(c)$) of the whole effective coupling between
the $Z^{(1)}$ ($Z^{\prime (1)}$) boson and fermions over the same coupling for
the $Z$ (would be $Z'$) boson zero mode, takes into account the wave function
overlap between the gauge boson KK excitations and the SM fermions.  The ratios
$Q(c)$ and $Q'(c)$ tend both to $\sqrt{2\pi kR_c}$ for $c \to -\infty$, while
one has $Q(c) \to -0.2$ and the other $Q'(c) \to 0$ in the limit $c \to +
\infty$; for the special value $c=0.5$ one has $Q(0.5) = 0$ and $Q'(0.5) 
\simeq 0.2$.\s 

\subsubsection*{2.3 Two scenarios for ${\rm U(1)_X}$} 
\label{sub:light}

In the present paper, we will discuss two possible scenarios for the abelian 
group  ${\rm U(1)_X}$. In a first scenario, that we will call RSa, we assume 
that ${\rm U(1)_{X}}\equiv {\rm U(1)_{B-L}}$, as in Ref.~\cite{ADMS}, so that 
the fermion representations/charges under the ${\rm SU(2)_L \times SU(2)_R 
\times U(1)_{B-L}}$ group are:
\begin{eqnarray}
Q_L^i &\equiv& ({\bf 2},{\bf 1})_{\frac{1}{6}}; \ \ \ 
u_R^i \in ({\bf 1},{\bf 2})_{\frac{1}{6}} \ \mbox{with} \ I_{3R}^{u^i_R}=+\frac{1}{2}; \
\nonumber \\
L_L^i &\equiv& ({\bf 2},{\bf 1})_{-\frac{1}{2}}; \
\ell_R^i \in ({\bf 1},{\bf 2})_{-\frac{1}{2}} \ \mbox{with} \ I_{3R}^{\ell^i_R}=-\frac{1}{2}
\end{eqnarray}
for the left--handed ${\rm SU(2)_L}$ doublets and the right--handed 
leptons and up--type quarks, and
\begin{eqnarray}
\mathbf{RSa}: \ \ d_R^i \in ({\bf 1},{\bf 2})_{\frac{1}{6}} \ \mbox{with} 
\ I_{3R}^{d^i_R}= -\frac{1}{2} 
\label{eq:scenA}
\end{eqnarray}
for the right--handed down--type quarks, $i=\{1,2,3\}$ being a generation 
index. Here, the Yukawa coupling terms for the quarks are written in terms of 
an invariant operator as, 
\begin{eqnarray}
\overline{({\bf 2},{\bf 1})}_{\frac{1}{6}}  ({\bf 2},{\bf 2})_{0}  
({\bf 1},{\bf 2})_{\frac{1}{6}} \label{eq:InvYuk}
\end{eqnarray}
the Higgs boson being embedded in a bidoublet of ${\rm SU(2)_L \times 
SU(2)_R}$.\s 

In the second scenario, that we will denote RSb, only the isospin assignment 
$I_{3R}$ [and thus the $Q_X$ charges as imposed by eq.~(\ref{eq:Ycond}), 
the SM hypercharge $Y$ being fixed] for the down--type right--handed quarks 
are modified with respect to the previous scenario:
\begin{eqnarray}
\mathbf{RSb}: \ \
d_R^i \in ({\bf 1},{\bf 2})_{-\frac{5}{6}} \ \mbox{with} \ I_{3R}^{d^i_R}=
+\frac{1}{2} 
\label{eq:scenB}
\end{eqnarray}
Here, we consider the possibility suggested recently in Ref.~\cite{AFB-Rold}
that the isospin up--type quarks $(u,c,t)$ acquire masses through a Yukawa
coupling of the type eq.~(\ref{eq:InvYuk}), whereas the down--type quarks
$(d,s,b)$ become massive via another operator. In general, this latter operator
will violate the custodial symmetry protecting the charges $Q_Z^{d^i}$, but it 
is natural to assume that its coefficient is small (in particular, in order to 
generate the small ratio $m_b/m_t$) so that the resulting $\Delta Q_Z^{d^i}/
Q_Z^{d^i}$ is also small.\s

As mentioned previously, the scenario RSa has been discussed in detail in
Ref.~\cite{ADMS}.  Because of the bulk custodial isospin gauge symmetry, an
acceptable fit of the ``oblique'' parameters \cite{PTparameters} $S$ and $T$
[including, after a redefinition, the coefficients $x$ and $y$ of fermion
operators for light fermions: $u,c,d,s, \ell^\pm,\nu$] can be achieved for
$M_{KK} \gtrsim 3$ TeV and $c_{\rm light}$ sufficiently larger than $0.5$,
irrespectively of the value of the coupling constant $g_{Z'}$ of the new $Z'$
boson, as shown in Ref.~\cite{ADMS} for the case ${\rm U(1)_{X}}={\rm
U(1)_{B-L}}$, that is, $Q_X=\frac12 (B-L)$ for all fields.  The heavy bottom
and top quarks must be treated separately since, in order to reproduce the
large value of the top quark mass through the geometrical mechanism developed
in the beginning of this section, one should typically have $c_{Q_L}<0.5$ 
[with $c_{Q_L}\!=\!c_{b_L}\!=\!c_{t_L}$ since $b_L$ and $t_L$ belong to the
same ${\rm SU(2)_L}$ multiplet] and $c_{t_R}<0.5$ \cite{HSquark} so that the
coefficients $x,y$ of operators for the $b$ and $t$ quarks cannot be redefined
into the $S$ and $T$ parameters \cite{ADMS}. Note that in order to maximize the
top quark mass, one can choose $c_{t_R} \sim -0.5$; a smaller value cannot be
taken otherwise the down--type component $b'_R$ of the ${\rm SU(2)_R}$ doublet
would have a too light KK excitation which is experimentally ruled
out\footnote{In the present analysis, we will assume that the ${\rm SU(2)_R}$
partners of the right handed quarks, in particular $b'_R$, are too heavy to
affect the electroweak precision data. This statement has to be quantified as
the mass of this new quark is related to the masses of the KK excitations of
gauge bosons.  For $b'_R$ states lighter than those assumed here, one would
need to include the fermionic $b^{(0)}$--$b_R^{\prime (1)}$ mixing which might
affect the $Zb \bar b$ vertex as is intensively discussed in
Refs.~\cite{Carena}.  Nevertheless, note that the induced contribution to
$\Delta (Q_Z^{b_L})^2/ (Q_Z^{b_L})^2$ can be compensated by $\Delta
(Q_Z^{b_R})^2/(Q_Z^{b_R})^2$ so that the resulting deviation in $R_b$ vanishes
(as will be seen later).  In a complete analysis, the mixing of both fermion
and gauge boson excitations might need to be taken into account.}.\s 

In the second scenario RSb, where the right--handed down--type quarks have
different $I_{3R}$ values from RSa,  the $Q_{Z'}$ charges of the quarks $d_R^i$
are different than previously. However, in the case of light quarks, 
$c_{\rm light}$ values larger than $0.5$ can always induce a decrease of $\vert
\Delta Q_Z^{d_R}/Q_Z^{d_R} \vert$ sufficient not to generate unacceptable
deviations in electroweak observables [the part independent of $F(c_{\rm
light})$ gives small contributions]? This is guaranteed by the condition
$M_{KK} \gtrsim 3$ TeV, exactly as in scenario RSa. As before the bottom and
top quarks, with $c<0.5$, must be treated separately.\s

The theoretical elements given in this section are sufficient to discuss the
effect of the RS scenario on the high precision measurements and show how the
$A_{FB}^b$ anomaly can be resolved, while keeping the other electroweak
observables in good agreement with experimental data.

\subsection*{3. Resolving the $\mathbf{A_{FB}^b(M_Z)}$ puzzle}

\subsubsection*{3.1 The FB asymmetry on the Z pole} 

On top of the $Z$ boson resonance, the forward--backward asymmetry for $b$ 
quarks, $A_{FB}^b (M_Z)$, can be written in the SM, in terms of the $Z$ 
charges to the left-- and right--handed chiralities introduced in the previous 
section, as  
\beq A_{FB}^b (M_Z)= \frac{3}{4}   \, 
\frac { (Q_Z^{e_L})^2 - (Q_Z^{e_R})^2 } { (Q_Z^{e_L})^2 + (Q_Z^{e_R})^2 } \, 
\frac { (Q_Z^{b_L})^2 - (Q_Z^{b_R})^2 } { (Q_Z^{b_L})^2 + (Q_Z^{b_R})^2 } \, 
\equiv \, \frac{3}{4} A_e \, A_b \, . 
\label{def:afb}
\eeq

As discussed earlier, it is the electroweak observable for which the deviation
between the experimentally measured value and the SM expectation is the highest
\cite{LEPex}: $A_{FB}^b=0.0992 \pm 0.0016$, while the SM fit gives
$A_{FB}^b=0.1037 \pm 0.0008$. This effect, nearly at the 3$\sigma$ level, could
be attributed to the special status of the $Zb\bar b$ vertex but this seems to
contradict the nice agreement observed for the partial width of the decay $Z\to
b\bar b$ which reads in the SM
\beq
R_b \equiv \frac{\Gamma (Z \to b\bar{b}) }{\Gamma (Z \to {\rm hadrons}) }
= \frac{ (Q_Z^{b_L})^2 + (Q_Z^{b_R})^2 } {\sum_{q} [ (Q_Z^{q_L})^2 + 
(Q_Z^{q_R})^2 ]} \ \  (q \neq t)
\label{def:rb}
\eeq
and is measured to be $R_b = 0.21629 \pm 0.00066$ while in the SM, one has
$R_b=0.2158$, meaning that the deviation is only at the $+0.7 \sigma$ level
\cite{LEPex}.\s 

In fact, if one uses the fitted value of $A_e$ deduced from leptonic data
assuming lepton universality, on obtains $A_b=0.882 \pm 0.017$ which is
$3\sigma$ away from the value $A_b=0.9347\pm 0.0001$ predicted in the SM. In
addition, if in the spirit of this paper one does not assume lepton
universality and uses the most precise result for $A_e$ obtained from the
measurement of the longitudinal polarization asymmetry $A_{LR}$ at SLC,
$A_e=0.1514 \pm 0.0022$, one has $A_b=0.874 \pm 0.019$ which is
$3.2\sigma$ away from the SM value. This  $(-6\pm 2.1) \sigma$ effect has to be
contrasted with the agreement observed on $R_b$ at the 0.3\% level. Note that
the direct measurement of $A_b$, using the FB  asymmetry with
polarized electrons at SLC, gives $A_b=0.925\pm 0.020$, consistent with the SM
but not precise enough to rule out the LEP1 value; combining both measurements
gives $A_b=0.900 \pm 0.013$  hence a $-3.86 \pm 1.44\%$ effect.\s

Using eqs.~(\ref{def:afb}) and (\ref{def:rb}) for the observables $A_{FB}^b$ 
and $R_b$, since the New Physics alters significantly only the 
left-- and right--handed $Zb\bar b$ couplings (the corrections $\Delta Q_Z^e
/Q_Z^e$ will turn out to be typically weaker as we must restrict to $c_{\rm 
light}$ larger than $0.5$), $Q_Z^{b_L}=-\frac{1}{2} + 
\frac13 s_W^2$ and $Q_Z^{b_R}= \frac{1}{3} s_W^2$ with $s_W^2 =\sin^2\theta_W 
\approx \frac14$, and noticing that
$(Q_Z^{b_L})^2/ (Q_Z^{b_R})^2 \sim 30$ which allows to safely use the
approximation $(Q_Z^{b_L})^2 \gg (Q_Z^{b_R})^2$, the cancellation of the new
effects in $R_b$ occurs if $\Delta (Q_Z^{b_R})^2 \simeq - \Delta
(Q_Z^{b_L})^2$ while the effect on the asymmetry would be in this case
\beq
\frac{ \Delta A_{FB}^b }{ A_{FB}^b } \simeq \frac{\Delta A_b } {A_b}
\simeq 2 \frac{ \Delta (Q_Z^{b_L})^2}{ (Q_Z^{b_L})^2} \simeq -2 \frac{
\Delta (Q_Z^{b_R})^2}{ (Q_Z^{b_L})^2}.
\eeq 
One then obtains the needed deviations  for the squared left-- and 
right--handed $Z b\bar b$ couplings to fully explain the $3\sigma$ anomaly
in $A_{FB}^b (M_Z)$  
\beq 
\Delta (Q_Z^{b_R})^2/ (Q_Z^{b_R})^2 \simeq (58 \pm 22)\% \ , \ \ \ 
\Delta (Q_Z^{b_L})^2/ (Q_Z^{b_L})^2 \simeq (-1.92 \pm 0.72)\%
\label{deviation+}
\eeq
where the uncertainties are due to the statistical errors.\s

Thus, as stated earlier, if New Physics has to explain the $A_{FB}^b$ anomaly 
while keeping $R_b$ SM--like, one needs a drastic change, at least $\sim 30\%$,
of the right--handed coupling $Q_Z^{b_R}$ and only a small change, at the
percent level, of the left--handed one. Moreover, the deviations in the  two
couplings squared should be  {\it opposite} in sign.  

\subsubsection*{3.2 The FB asymmetry in scenario RSa}

In the scenario RSa in which the abelian group is ${\rm U(1)_{X}}={\rm 
U(1)_{B-L}}$ as in Ref.~\cite{ADMS}, the new contributions to the $Zb\bar b$
coupling, using the charges defined earlier and assuming $\sin^2 \theta'
\ll 1$ are
\begin{eqnarray}
\frac{\Delta Q_Z^{b_L} } {Q_Z^{b_L}} \simeq \frac{M_Z^2}{(0.4 M_{KK})^2}
\bigg [ 1 + \frac{1}{4F(c_{b_L})} - 0.09 \bigg ] F(c_{b_L}) \, , \\
\frac{\Delta Q_Z^{b_R}} {Q_Z^{b_R}} \simeq \frac{M_Z^2}{(0.4 M_{KK})^2}
\bigg [ 1 + \frac{1}{4F(c_{b_R})} + \frac{1.5}{\sin^2\theta'}\bigg ] F(c_{b_R}) 
\label{DeltaQapp}
\end{eqnarray}

Thus, for moderate KK masses, $M_{KK} \sim 3$ TeV, and small values of the new
mixing angle $\sin \theta'$ [but not too small: $\sin \theta' \gtrsim 0.1$, so
that the coupling $g_{Z'} = \tilde g'/\sin \theta'$ remains perturbative], one
can generate a large correction $\Delta Q_Z^{b_R}/Q_Z^{b_R}$, while $\Delta
Q_Z^{b_L}/Q_Z^{b_L}$ remains small. A significant hierarchy between $\Delta
Q_Z^{b_R}/Q_Z^{b_R}$ and $\Delta Q_Z^{b_L}/Q_Z^{b_L}$ can also be generated
through the $F(c_{b_{L,R}})$ functions. However, since the latter function is
always negative, the two contributions $\Delta Q_Z^{b_L}/Q_Z^{b_L}$ and $\Delta
Q_Z^{b_R}/Q_Z^{b_R}$ have always {\it the same sign} (negative) and this does
not allow to cure the $A_{FB}^b$ anomaly while leaving $R_b$ almost unaffected,
eq.~(\ref{deviation+}). \s

A solution to this problem, as pointed out in Ref.~\cite{AFB-mix} in a 
different context,  would be to generate an even larger correction
$\Delta Q_Z^{b_R}$ to {\it flip the sign} of the overall $Z b_R\bar b_R$ charge
$Q_Z^{b_R}+\Delta Q_Z^{b_R}$ since in the SM, the charge squared $(Q_Z^{b_R})^2
\sim \frac{1}{30}$ is naturally the smallest one. The needed deviations would
be in this case, 
\beq \Delta Q_Z^{b_R} / Q_Z^{b_R} \simeq (-230 \pm 10)\% \ , \
\ \ \Delta Q_Z^{b_L} / Q_Z^{b_L} \simeq (-1 \pm 0.4)\% \label{deviation-} 
\eeq

\begin{figure}[!h]
\vspace*{-2.cm}
\begin{center}
\epsfig{file=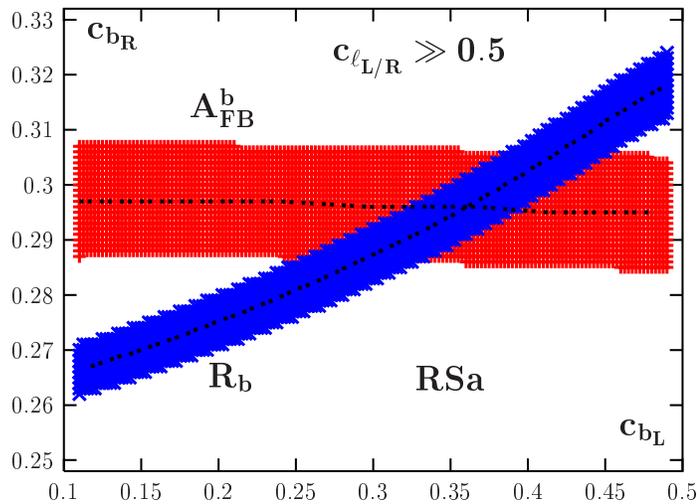,width=16.cm} 
\end{center}
\vspace*{-14.2cm}
\caption{Contour plots in the plane $[c_{b_L}, c_{b_R}]$ for $M_{KK}=3$ TeV and 
$\sin\theta'=0.1$, for which $A_{FB}^b$ and $R_b$ in the scenario RSa are 
equal to their experimental values (dotted lines) and are within their $\pm 1
\sigma$ bands; $c_{\rm light} \gg 0.5$ for leptons and light quarks.} 
\label{ContourPlot1}
\vspace*{-.1cm}
\end{figure}

In Fig.~\ref{ContourPlot1}, we show the values in the $[c_{b_L},c_{b_R}]$
parameter space for which the predicted values of $A_{FB}^b$ and $R_b$ in this
RS scenario are equal and within $\pm 1\sigma$ to the experimental
measurements. The other inputs are $M_{KK}\!=3\!$ TeV [which is allowed by the
oblique $S,T$ parameters] and $\sin \theta'\! =\!0.1$
[i.e.  close to the lowest value allowed for the perturbativity of the $g_{Z'}$
coupling].  As can be seen, there is a set of $[c_{b_L},c_{b_R}]$ assignments
for which $A_{FB}^b$ and $R_b$ are equal to their experimental values. Indeed,
for $c_{b_L}=0.375$ and $c_{b_R}=0.296$, with again  $c_{\ell_L} = c_{\ell_R}
\gg 0.5$, one finds for the two observables $A_{FB}^b = 0.0992$ and
$R_b=0.21629$, $i.e.$ the exact experimental values.\s

For heavier KK excitations $M_{KK}>3$ TeV and fixed $\sin\theta'$, lower 
$c_{b_R}$ and $c_{b_L}$ values are required to increase $|F(c)|$ and thus to 
compensate the $1/M_{KK}^2$ decrease of $|\Delta Q_Z^{b_{L,R}}/Q_Z^{b_{L,R}}|$,
eq.~(\ref{Z'charges}).  Alternatively, for $\sin\theta'$ larger than 0.1, lower
$c(b_R)$ values which compensate the decrease of the $g_{Z'}$ effect in $\Delta
Q_Z^{b_R}/ Q_Z^{b_R}$ are required [the $g_{Z'}$ coupling is annihilated by the
charge $Q_{Z'}^{b_L}$ in $\Delta Q_Z^{b_L}/Q_Z^{b_L}$]. \s

Thus, it is remarkable that {\it both} the measurements of $A_{FB}^b$ and $R_b$
can be reproduced  for $c_{b_{L/R}}$ values of order unity, which means that no
additional scale is introduced in the model as explained in section 2.1. Given
the above discussion on $c$ parameters, the favored values for $M_{KK}$ and
$\sin \theta'$ (the $kR_c$ influence being weak), with regard to a realistic
heavy quark mass spectrum, are those chosen for Fig.~\ref{ContourPlot1}. For
the optimal $c$ values, $c_{b_L}=0.47$ and $c_{b_R}=0.305$ which reproduce the
experimental values of $A_{FB}^b$ and $R_b$ [as noted earlier, the value
$c_{t_R}=-0.5$ can be chosen as to maximize $m_t$, while $c_{t_L} =c_{b_L}$],
the approximate $m_t$ and $m_b$ values which can be obtained via the
geometrical mechanism discussed in section 2, are such that $m_b \sim 40$ GeV
and $m_t \sim 75$ GeV. Both of these generated masses are far from the
experimentally measured values. However, these differences can easily originate
from the full three--flavor mixing effects\footnote{Note that within the
full three--flavor approach required to give valid conclusions on the exact
quark/lepton mass spectrum, one should also consider the experimental bounds on
FCNC processes which translate into a lower bound on $M_{KK}$
\cite{HSquark,FCsign}. It was shown recently \cite{HSquark} that this bound can
be softened down to $M_{KK} \gtrsim 1$ TeV for certain geometrical
configurations reproducing the fermions masses.}, significant contribution from
the mixing with the first KK excitations as discussed in Ref.~\cite{Victoria}
in the case of the top quark, or from a reasonably small hierarchy among the
five--dimensional Yukawa couplings.  Indeed, the quark mass values given above
correspond to a theory in which the Higgs boson is exactly confined on the
TeV-brane with the natural choice $Y^{(5)}=\kappa / \sqrt{k}$, where $\kappa$
(relating the five--dimensional Yukawa coupling to the parameter $k$ as usual
\cite{RSloc,Victoria,HSquark}) taken equal to unity.  The actual top and bottom
masses are reproduced for $\kappa_{t} \sim 2$ and $\kappa_{b} \sim
0.1$; such $\kappa$ values are acceptable as they correspond to almost
universal $Y^{(5)}$ couplings and introduce mass parameters of an order of
magnitude relatively close to the fundamental energy scale $k$.\s

Finally, let us make a few remarks on the case of leptons $\ell = e,\mu$ and
$\tau$. The constraints from LEP1 and SLC measurements are very tight since the
decay widths $\Gamma_\ell$ and the asymmetries $A_\ell$ and, in particular the
measurement of $A_e$ from the left--right asymmetry and $A_\tau$ in polarized
$\tau$ decays, are determined at the percent level.  However, it is known that
$A_e$, which is the most precisely measured quantity, shows a $\sim 1.7$
standard deviation with respect to the prediction in the SM, $A_e^{\rm
SLD}=0.1516\pm 0.0021$ compared to the fitted value $A_e^{\rm SM}=0.1480$ [the
deviation drops slightly in $A_\ell$ when all leptonic asymmetries are averaged
and lepton--universality is assumed].  It is tempting to try to explain this
discrepancy within the RS framework with the extended gauge structure. To do
so, one can choose a value $c_{e_R}\sim 0.56$ [in agreement with the analysis
of Ref.~\cite{ADMS} of oblique $S$ and $T$ parameters] while the value of
$c_{e_L}$ can be taken to be much larger, $c_{e_L}\gg 0.5$.  Again for $M_{KK}
=3$ TeV and $\sin\theta'=0.1$, one obtains then for the electron asymmetry
$A_e=0.1504$, which is only $-0.5\sigma$ away from the measured value by SLD.
This choice of parameters will also generate an additional deviation in
$A_{FB}^b \equiv \frac34 A_e A_b$; still, there is a choice of $c$ parameters
in the $b$ quark sector which keeps the FB asymmetry close to its measured 
value. Thus, a judicious choice of the $c$ parameters in both the bottom quark
and electron sectors can allow to relax the  tension between the two
measurements which provide the best individual determinations of the weak
mixing angle $\sin^2\theta_W$ and, thus, provide a very nice global fit to the
precision data.\s

\subsubsection*{3.3 The FB asymmetry in scenario RSb}

In the second scenario RSb, where only the $I_{3R}$ values and thus the $Q_X$ 
charges for the down--type right--handed quarks are modified with respect to 
the previous scenario in which $Q_X=\frac12(B-L)$, the contributions of the
KK excitations to the $Zb\bar b$ coupling are 
\begin{eqnarray}
\frac{\Delta Q_Z^{b_L} } {Q_Z^{b_L}} \simeq \frac{M_Z^2}{(0.4 M_{KK})^2}
\bigg [ 1 + \frac{1}{4F(c_{b_L})} - 0.09 \bigg ] F(c_{b_L}) \, , \\
\frac{\Delta Q_Z^{b_R}} {Q_Z^{b_R}} \simeq \frac{M_Z^2}{(0.4 M_{KK})^2}
\bigg [ 1 + \frac{1}{4F(c_{b_R})} - \frac{1.5}{\sin^2\theta'}\bigg ] F(c_{b_R}) 
\label{DeltaQappBIS}
\end{eqnarray}
where we have used again the approximation $\sin^2 \theta' \ll 1$. The effect
of the isospin choice $I_{3R}^{b_R}=+\frac12$ will be simply to reverse the
sign of the contribution proportional to $\sin^{-2}\theta'$ in the charge
$Q_Z^{b_R}$ and, therefore, one can have contributions $\Delta Q_Z^{b_L}/
Q_Z^{b_L} <0$ and $\Delta Q_Z^{b_R}/Q_Z^{b_R}>0$ without flipping the sign of 
the charge $Q_Z^{b_R}$ in contrast to the previous situation. One can then cure
the $A_{FB}^b(M_Z)$ anomaly and simultaneously leave $R_b$ almost unchanged, 
eq.~(\ref{deviation+}), without too large corrections for the $b_R$ coupling 
to the $Z$ boson. \s

\begin{figure}[!h]
\vspace*{-2.3cm}
\begin{center}
\epsfig{file=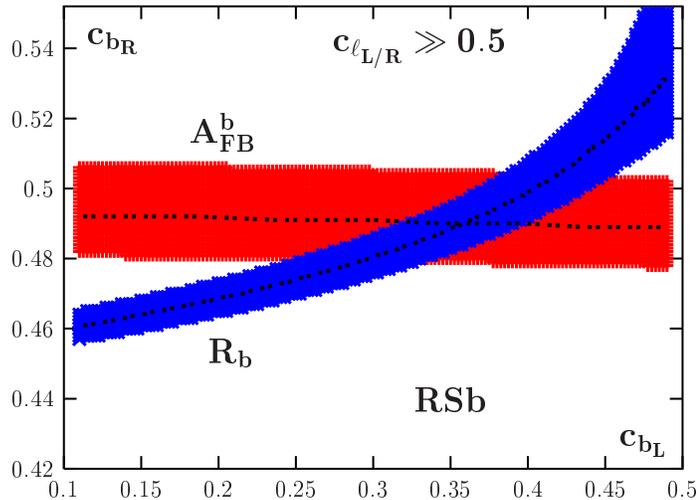,width=16.cm} 
\end{center}
\vspace*{-14.3cm}
\caption{Contour plots in the plane $[c_{b_L}, c_{b_R}]$ for $M_{KK}=3$ TeV and 
$\sin\theta'=0.1$, for which $A_{FB}^b$ and $R_b$ in the scenario RSb are 
equal to their experimental values (dotted lines) and are within their $\pm 1
\sigma$ bands; $c_{\rm light} \gg 0.5$ for leptons and light quarks.} 
\label{ContourPlot2}
\vspace*{-.4cm}
\end{figure}

In Fig.~\ref{ContourPlot2}, assuming $M_{KK}=3$ TeV (in order to address the
gauge hierarchy problem) and $\sin\theta'=0.1$ and using $c_{\rm light} \gg 0.5$
in such a way that the observables involving leptons and light quarks are not
affected, we show in the $[c_{b_L},c_{b_R}]$ plane the predicted values of
$A_{FB}^b(M_Z)$ and $R_b$ in this second scenario (the dotted lines); the bands
represent the two observables when the $1\sigma$ experimental errors are
included. One can see that, as previously, there is a set of $[c_{b_L},c_{b_R}]$
assignments for which the experimental values of $A_{FB}^b$ and $R_b$ are
reproduced. However, here, one needs values of  $c_{b_R}$ very close to $0.5$ to
resolve the $A_{FB}^b(M_Z)$ anomaly. Note also that, exactly as in scenario RSa,
with a judicious choice of $c$  parameters for the electron, one can also
explain the slight discrepancy of  $A_e$ with the value measured at SLD.

\subsection*{4. $\mathbf{A_{FB}^b}$ outside the Z-pole}

To derive the FB asymmetry [and the production cross section] of the bottom
quarks outside the Z--pole, one has to take into account the usual
$s$--channel  virtual photon and $Z$ boson exchanges but also, in principle, 
the exchange of the heavy $KK$ states $V^{(1)}\equiv \gamma^{(1)}, Z^{(1)}, 
Z^{\prime (1)}$ as well as their interference terms. At the tree level, the 
differential cross for the production of a fermion pair in $\ee$ collisions, 
$\ee \to f\bar f$, can be written as 
\begin{eqnarray}
\frac{{\rm d} \sigma_f}{{\rm d} \cos \theta } = \frac{3}{8} \ \sigma_0
\ N_c \beta_f \left[ (1+\beta_f^2 \cos^2 \theta)Q_1+ (1-\beta^2_f)Q_2
+2  \beta_f  \cos \theta Q_3 \right]
\label{diffdis}
\end{eqnarray}
\nn with $\beta_f=(1-4m_f^2/s)^{1/2}$ the velocity of the final state fermion,
$\theta$ specifying its direction with respect to the incoming electron and
$N_c$ its color factor and where $\sigma_0=4\pi \alpha^2/3s$ is the point--like 
QED cross section for muon pair production. Integrating over the scatting 
angle, the total production cross section $\sigma_f(s)$ and the FB
asymmetry $A_{FB}^f(s)$ are then given by
\begin{eqnarray}
\sigma_f = &  \frac{3}{4} \sigma_0 N_{c} \beta_f
\left[(1+\frac{1}{3}\beta_f^{2}
) Q_{1}+(1-\beta_f^{2})Q_{2} \right]  \; \; & \stackrel{ \sqrt{s} \gg m_{f}}
{\longrightarrow} \; \; N_{c} \sigma_0 Q_{1} \\
A_{FB}^f = & \beta_f Q_{3} \left[ (1+\frac{1}{3}\beta_f^{2})Q_{1}+
(1-\beta_f^{2})
Q_{2} \right]^{-1} \; \; & \stackrel{ \sqrt{s} \gg m_{f}} {\longrightarrow}
\; \; \frac{3}{4} \frac{Q_{3}}{Q_{1}} 
\end{eqnarray}
In terms of the helicity amplitudes $Q_{ij}$ with $i,j=L,R$, the charges 
$Q_{1,2,3}$ are given by \cite{Charges}
\begin{eqnarray}
Q_{1} &=& \frac{1}{4} \left[ \; |Q_{LL}|^{2} + |Q_{RR}|^{2} +
|Q_{RL}|^{2} +|Q_{LR}|^{2} \right]  \nonumber \\
Q_{2} &=& \frac{1}{2} {\rm Re} \ \left[ \; Q_{LL}Q_{RL}^{*} +
Q_{RR}Q_{LR}^{*} \right]  \nonumber \\
Q_{3} &=& \frac{1}{4} \left[ \; |Q_{LL}|^{2} + |Q_{RR}|^{2} -
|Q_{RL}|^{2} - |Q_{LR}|^{2} \right]
\end{eqnarray}
Taking into account all exchanged  gauge bosons, the helicity amplitudes 
$Q_{ij}$ read 
\begin{equation}
Q_{ij} = Q^e_\gamma  Q^f_\gamma + \frac{Q_Z^{e_i}Q_Z^{f_j}} {s_W^2c_W^2}
\frac{s}{s- M_Z^2+i\Gamma_Z M_Z} +\sum_{V}
\frac{ g_V^2 } {e^2} Q_V^{e_i} Q_V^{f_j} Q_V(c_{e_i}) Q_V(c_{f_j}) 
\frac{s}{s- M_V^2}
\label{eq:Qij} 
\end{equation}
where $g_V=e,g_{Z}=e/s_Wc_W,g_{Z'}$ for $V=\gamma^{(1)},Z^{(1)},Z^{\prime(1)}$,
$Q_{ \gamma^{(1)}, Z^{(1)}}=Q(c)$ and $Q_{ Z^{\prime (1)} }=Q'(c)$.\s
 
To evaluate the FB asymmetry for $b$--quarks just below or above the $Z$
resonance, $\sqrt s \sim M_Z$, one can safely neglect the contributions of the
KK excitations that are exchanged in the $s$--channel since these states are
relatively heavy, $M_{KK} \gtrsim 3$ TeV, and their couplings to the initial
state electrons are rather small. In fact, even at LEP2 energies, $\sqrt s \sim
200$ GeV, the effects of the exchanged heavy excitations  can be also neglected
as the experimental errors in the measurement of $A_{FB}^b (s)$ are rather large
\cite{AFB-above}.  One then needs to consider solely the photon and the $Z$
boson exchange contributions and include the mixing effect of the $Z$ boson with
the gauge boson KK states as discussed earlier. Note that one has to take into
account the mixing between the $B_0$ and $\bar B_0$ mesons when the $b$--quark
is tagged through its semi--leptonic decays for instance as is the case at
experiments far below the $Z$--pole from PEP to TRISTAN; the bare asymmetry has
to be then multiplied by a mixing factor $ A_{FB}^\ell =(1-2\chi) A_{FB}^b$
where the parameter $\chi$ is measured to be $\chi \simeq 0.125$ \cite{PDG}.\s

In principle, to have a very accurate determination of the asymmetry, some
higher order effects \cite{AFB-SM} need to be included in eq.~(\ref{diffdis}). 
Besides the $b$--mass effects and the QCD radiative corrections [involving again
the $m_b$ effects] which can be readily implemented \cite{AFB-SM,QCD-eff}, very
important effects are the electroweak radiative  corrections and, in particular,
initial state radiation (ISR) of photons which are crucial especially at
energies $\sqrt{s} \gsim M_Z$,  as it allows for the return to the $Z$--pole
where the $\ee\to b\bar b$ rate becomes huge, if no cut is applied on the photon
energy. The latter corrections can only be implemented in a Monte--Carlo based
approach which involves a full treatment of the kinematics.\s 

At the $Z$--pole, these higher order effects have been implicitly taken into
account as we have evaluated small deviations with respect to the SM prediction
given in Ref.~\cite{LEPex} and in which they are already incorporated. A few GeV
above and below the $Z$ pole, where high precision measurements have also been
performed by the LEP experiments \cite{LEPex}, this approximation might also
hold, except for initial state radiation which needs to be explicitly
implemented.  We have evaluated the FB asymmetry for $b$-quarks at the three
LEP1 running energies $\sqrt s=89.55, \, 91.26$ and 92.94 GeV in the SM and in 
the two scenarios RSa and RSb, including the $Z$ and photon exchange and their
interference as well as ${\cal O}(\alpha)$ initial state radiation. A constant
fudge factor close to unity allows to reproduce exactly the SM curves given in
Ref.~\cite{LEPex}, with the Higgs boson mass taken to be $M_H=100$ GeV. 
Assuming the same fudge factor, the variation of $A_{FB}^b$ with the c.m. energy
around the $Z$ pole in the two scenarios RSa [with $c_{b_L}=0.37,  c_{b_R}=0.287$
and $c_{\rm light} \gg 0.5$] and RSb [with $c_{b_L}=0.36,  c_{b_R}=0.49$ and
$c_{\rm light} \gg 0.5$] which reproduce as closely as possible the experimental
values of $A_{FB}^b (s)$ and $R_b$ (the fit depends only mildly on $M_{KK}$ and
$\sin\theta'$), is shown in the  left--hand side of Fig.~3 and is compared to
the one in the SM; see also Tab.\,1. For these sets of parameters  corresponding
to the best global fit,  we find: $R_b=0.21618$, $\Delta
Q_Z^{b_L}/Q_Z^{b_L}=-0.79\%$,  $\Delta Q_Z^{b_R}/Q_Z^{b_R}=-227\%$ [RSa] and
$R_b=0.21629$,  $\Delta Q_Z^{b_L}/Q_Z^{b_L}=-0.85\%$, $\Delta
Q_Z^{b_R}/Q_Z^{b_R}=+27\%$ [RSb].\s

For $\sqrt s=91.26$ GeV, i.e. very close to the $Z$ pole, the experimental
value for $A_{FB}^b(s)$ lies almost on top of the predictions in the two RS
scenarios. The measurement of the asymmetry at $\sqrt s=89.55$
GeV is also perfectly reproduced in the RSb scenario, contrary to the SM case
in which the prediction is about $1.2\sigma$ away and to RSa  for which there is
a large deviation, $2.8\sigma$. In turn, the point at $\sqrt s=92.94$ GeV is
perfectly reproduced only in the RSa scenario: the prediction in the SM is
$2.7\sigma$ away, while in the scenario RSb, the deviation from the
experimental central value is at $+2\sigma$ level which is acceptable,
though.  If one adds the three measurements around the $Z$ pole, the $\chi^2$
is much better in the RSb model, $\chi^2_{\rm RSb} \simeq 4.2$, than in the SM,
$\chi^2_{\rm SM} \simeq 15.3$. In the scenario RSa, the  $\chi^2$ is 
better than in the SM, $\chi^2_{\rm RSa} \simeq 8.5$, despite of the large
deviation at $\sqrt s =89.55$ GeV. This is a consequence of the fact that the
point at $\sqrt s \simeq M_Z$ has a much larger weight in the averaging
procedure as its corresponding experimental error is much smaller\footnote{One
should note that, in fact, the measurements at the two energies $\sqrt s=89.55$
and 92.94 GeV are included in the averaged value of $A_{FB}^b$ given by the LEP
experiments \cite{LEPex}.}. \s

\begin{figure}[!h]
\vspace*{-2.6cm}
\hspace*{-1.5cm}
\epsfig{file=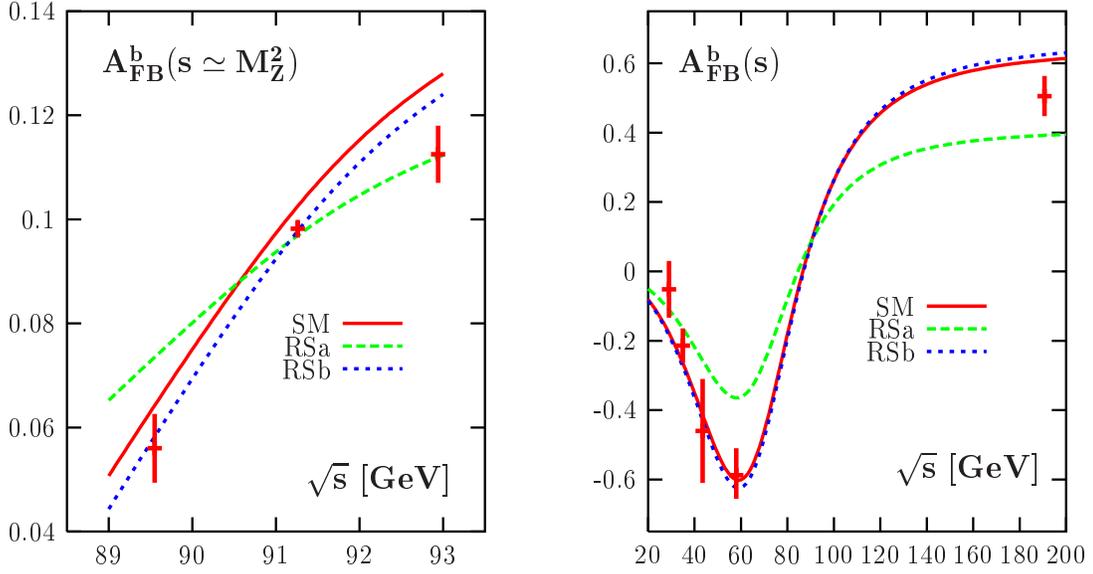,width=18.cm} 
\label{Afb-plot}
\vspace*{-16.cm}
\caption{The FB asymmetry $A_{FB}^b$ at LEP1 (left) and outside the $Z$ 
resonance (right) as a function of the c.m. energy. Shown are the predictions 
in the SM and in the two RS scenarios RSa [with $c_{b_L}=0.37, c_{b_R}=0.287, 
c_{\rm light} \gg 0.5$] and RSb [with $c_{b_L}=0.36, c_{b_R}=0.49, 
c_{\rm light} \gg 0.5$] for $M_{KK}=3$ TeV and $\sin\theta'=0.1$,  as well as 
the various experimental measurements with their error bars.}
\label{Afbplot}
\end{figure}

\begin{table}[!h] 
\vspace*{.1cm}
\begin{center} 
\renewcommand{\arraystretch}{1.2}
\begin{tabular}{|c|c||cc|cc|cc|} \hline
$\sqrt s$ & Experimental value &  SM  sd & $\chi^2$ &  RSa sd & $\chi^2$ & RSb sd & $\chi^2$ \\ \hline
89.55 GeV & $0.0560\pm0.0066$ &  $+1.2$   &$$ &      $ +2.8$ & $$ &       $+0.4$  &$$ \\  
91.26 GeV & $0.0982\pm0.0017$ &  $+2.5$   &$15.3$ &  $+0.5$  & $8.5$ &    $-0.1$  &$4.2$ \\
92.94 GeV & $0.1125\pm0.0055$ &  $+2.7$   &$$ &      $+0.4$  & $$ &       $+2.0$  &$$ \\ \hline 
29 GeV & $-0.052 \pm 0.081 $   & $-1.5$   &$$ &      $-0.8 $  &$$ &       $-1.6$  &$$ \\
35 GeV & $-0.214 \pm 0.050 $   & $-1.0$   &$$ &      $+0.9 $  &$$ &       $-1.2$  &$$ \\
44 GeV & $-0.460 \pm 0.147 $   & $+0.3$   &$7.6$ &   $+1.3 $  &$14$ &     $+0.1$  &$9.8$ \\
58 GeV & $-0.588 \pm 0.078 $   & $-0.2$   &$$ &      $+2.7 $  &$$ &       $-0.4$  &$$ \\
190.7 GeV& $0.5055 \pm 0.058$  & $+2.0$   &$$ &      $-1.9 $  &$$ &       $+2.3$  &$$ \\ \hline
\end{tabular} 
\caption{The asymmetry $A_{FB}^b(s)$ [corrected for $B$ meson mixing] as 
measured at various c.m. energies, and the standard deviations in the SM and 
in the two scenarios RSa and RSb. The measured values and the errors at LEP 
energies are from Refs.~\cite{LEPex,AFB-above} and at lower energies, they are 
taken from the summary given in  Ref.~\cite{Klaus}.} 
\end{center}
\vspace*{-.7cm}
\end{table}

Let us now turn to energies far below [from PEP to TRISTAN] and far above 
[LEP2] the $Z$--pole for which the experimental values of $A_{FB}^b$ with the 
error bars are given, respectively, in Refs.~\cite{AFB-below,AFB-mix} and 
Ref.~\cite{AFB-above}; these values are summarized in the lower part of Tab.~1.
We have calculated the asymmetry $A_{FB}^b (s)$ as a function of the c.m. 
energy, taking into account only the effects of $B_0$--$\bar B_0$ mixing.  The
results for the FB asymmetry in the SM and the two RS scenarios are shown in the
right--hand side of Fig.~\ref{Afbplot} and are compared to the experimental
values with the error bars that are given in Table 1.\s 

Here, although the experimental errors are rather large, the general trend is
clearly that the predictions for $A_{FB}^b(s)$ in the RSb scenario as well as in
the SM [which are almost identical] are closer to the measured values.  There
are large deviations between the predictions in scenario RSa and the
experimental values, in particular at the energy $\sqrt s =58$ GeV, where the
difference is at the level of $2.7\sigma$. Adding all the measurements outside
the $Z$--pole region,  the $\chi^2$ in the SM and in the RSb scenarios,
$\chi^2_{\rm SM} \simeq 7.6$ and $\chi^2_{\rm RSb} \simeq 9.8$, are much better than 
in the RSa scenario, $\chi^2_{\rm RSa} \simeq 14$. \s 

In total, when all measurements around and off the $Z$ pole are considered and
if the measurement of $R_b$ is included, the agreement between the theoretical
prediction and the experimental data is excellent in the RSb model where the
overall $\chi^2$ is $\chi^2_{\rm RSb} \simeq 14$, which corresponds to a
probability\footnote{To assess the statistical significance of this result, one
should take into account the number of free parameters in the model. While
apparently $A_{FB}^b$ is adjusted by varying $c_{b_R}, M_{KK}$ and 
$\sin\theta'$, these parameters are highly correlated since they intervene
through the combination $F(c_{b_R})/(M_{KK} \sin\theta')^2$. There is,
therefore, effectively only one free parameter, since $F(c_L)$ has no influence
on $A_{FB}^b$ but is only used to tune $R_b$.} of a few percent.  In contrast,
the  prediction in scenario RSa is not satisfactory as  $\chi^2_{\rm  RSa}
\simeq 22.5$ (which corresponds to a probability that is one order of magnitude
worse than in RSb)  is as bad as in the SM, $\chi^2_{\rm SM} \simeq 23.4$. 
Thus, one can conclude that only in scenario RSb the  $A_{FB}^b$ anomaly is
resolved both at the $Z$--pole and outside the pole. We should however stress
again that, while our results for $A_{FB}^b$ on the  $Z$--pole, and to a lesser
extent a few GeV around the pole,  are quite robust  as we analyzed small
deviations compared to the SM values which include all  refinements and higher
order effects, the results off the $Z$--pole are less accurate. A full
Monte-Carlo analysis, including eventually the experimental  cuts [in particular
for ISR] is required to make very precise statements. \s

Before closing this section, let us make a brief comment on a scenario proposed
in Ref.~\cite{AFB-Rold} where, in addition to the possible modification of the
${\rm SU(2)_R}$ assignment of the $b_R$ quarks as in our scenario RSb, the
assignments for the left--handed quarks are also changed, and one has the
representation $Q_L^i \equiv ({\bf 2},{\bf 2})_{2/3}$. In this case, the two
multiplet embeddings for the isospin up--type $u_R^i$ quarks allowing invariant
Yukawa couplings are: $u_R^i \in ({\bf 1},{\bf 1})_{2/3}$ or $({\bf 1},{\bf
3})_{2/3} \oplus  ({\bf 3},{\bf 1})_{2/3}$. In this context, the necessary shift
in the $Z\bar b_R b_R$ coupling ($\Delta Q_Z^{b_R} / Q_Z^{b_R}>0$) that could
explain the $A_{FB}^b$ anomaly at the $Z$ pole can be achieved, assuming that
the smallness of the ratio $m_b/m_t$ comes from a strong hierarchy among
coupling constants \cite{AFB-Rold}.
However, we find that the best fit of $A_{FB}^b(s)$ and $R_b$ corresponds to
$\chi^2 \simeq 19.4$ which improves the SM case, but is not as good as our
scenario RSb. This fit is obtained for $c_{b_R}=0.525$  implying $\Delta
Q_Z^{b_R}/Q_Z^{b_R}=+8.5\%$ if $M_{KK} = 3$ TeV and $\sin \theta' \! = \! 0.1$,
the dependence on $M_{KK}$ and $\sin \theta'$ being small once again. The reason
for this degradation of the fit  is that in order to reproduce simultaneously
$R_b$ and $A_{FB}^b(s)$, one should have a correction $|\Delta Q_Z^{b_L} /
Q_Z^{b_L}|$ of order of the percent [{\it c.f.} eq.~(\ref{deviation+})] while in
this case, in contrast, the bidoublet nature of $b_L$ forces this correction to
vanish. The breaking due to the boundary conditions makes this vanishing
imperfect but these effects lead typically to too small deviations: $|\Delta
Q_Z^{b_L} / Q_Z^{b_L}| < 10^{-3}$ \cite{Carena}.

\subsection*{5. Expectations for the ILC}

In this section, we analyze the prospects for observing the effects of the new 
KK states predicted in the RS scenarios discussed here at the ILC, as it is  a
straightforward generalization of what has been discussed in the previous
sections. At the hadron colliders Tevatron and LHC, the situation is more
delicate and the search will be more complicated as the  main decay modes of
the KK states [contrary to conventional $Z'$ bosons from GUTs for which BR$( Z'
\to e^+ e^-, \mu^+\mu^-)$ is at the level of a few percent] will be into top and
bottom quark pairs which have large QCD backgrounds; see for instance the
general analyses performed in Ref.~\cite{KK-pp}. A detailed account of the
expectations at these these colliders in the particular model discussion here
will be given  in Ref.~\cite{DMR2}.

\subsubsection*{5.1 Expectations for bottom quarks} 

At the GigaZ option of the ILC, i.e. when running the collider at c.m. energies
close to the $Z$--pole, the LEP1 and SLC precision measurements discussed
previously can be vastly improved, as the planed luminosity will be much higher
and the longitudinal polarization of the initial beams available. This will
allow to have a more significant determination of the $Z b\bar b$ vertex and
thus, to probe the mixing between the $Z$ boson and the KK states with a much
higher accuracy.  The accuracies for determining $R_b$ and, using the
polarization asymmetry $A_e$ from $A_{LR}$ and $A_b$ from $A_{LR,FB}^b$ by
simply extrapolating the results from LEP1 and SLC are shown in Table 2.  For
the normalized partial width $R_b$, we assume that the double $b$--tag method
will be applied as for LEP1 but that, for a given purity, the $b$--jet tagging
efficiency will be multiplied by a factor of three as can be inferred from SLD.
For $A_b$, we use the effective efficiency from SLD multiplied by a factor of
two given the improved algorithms already foreseen and the better angular
coverage of the micro--vertex detector at the ILC. Finally, for $A_e$, we have
assumed that the dominant systematical error from SLD, beam polarization, will
be removed by using positron polarization. \s

In view of the very high sensitivity of GigaZ, one can stand a reduction of the
new effects predicted in the RS scenario by up to two orders of magnitude. This
would allow to set an upper limit on the KK mass in the 15--30 TeV range. As
pointed out already in Ref.~\cite{Hewett}, GigaZ through high precision
measurements, can therefore already fully cover the RS scenario in a range of
parameters which does not reintroduce fine--tuning. It can confirm without
ambiguity the indication observed on $A_e$ at SLD and establish at a very
sensitive level that light fermions are weakly affected, ruling out alternative
interpretations [note that in Little Higgs and Technicolor models, for
instance, there would be no large effects in the $b$--quark sector anyway]. The
measurement of $A_e$ gives the electron coupling to the $Z'$ boson and,
therefore, allows to estimate the interference term at higher c.m.  energy
between the $Z$ and photon exchanges and the KK excitations.

\begin{table}[!h] 
\vspace*{-.2cm} 
\begin{center}
\renewcommand{\arraystretch}{1.2} 
\begin{tabular}{|c||c|c|c|} \hline 
Energy  & Observables  & ~~~Error \%  ~~~& Deviation \%  \\ \hline 
GigaZ &  $R_{\rm b}$ & 0.015 & 0.3$\pm$0.3    \\ 
&  $A_{\rm b}$ & 0.03 & --3.9$\pm$1.4  \\ 
&  $A_{\rm e}$ & 0.01 & 2.4$\pm$1.4 \\ \hline 
500 GeV &  $A_{\rm LR}$ & 0.3 & +32  \\  \hline
\end{tabular}
\caption{Deviations and accuracies expected at GigaZ in the $b$ and electron 
sectors assuming the full accuracy provided by polarized electrons and 
positrons. For ILC operating at $\sqrt s=500$ GeV, one assumes a luminosity of 
500 fb$^{-1}$ with polarized electrons.} 
\end{center} 
\vspace*{-1.cm} 
\end{table} 

At higher energies, using the formulae derived in section 4 and assuming 
$c(e_R) =0.56$, one can easily predict the signals which can be observed at ILC.
For the $b$--tagging efficiency/purity at $\sqrt s=500$ GeV, we have assumed the
same efficiency for double tagging. In addition to the selection effect, one
should worry about the determination of the integrated luminosity and, for some
part of the analysis, the knowledge of the beam polarization. For  the
luminosity, ILC intends to achieve the 10$^{-4}$ level, which is sufficient to
fulfill our goals.  For the polarization, one can use the $\ee \to W^+W^-$
channel which has a strong polarization asymmetry, as a very effective and
realistic analyzer of electron polarization. Given the huge cross section of 
the process, there is no  limitation coming from statistics.\s

\begin{figure}[!h]
\vspace*{-.5cm}
\begin{center}
\epsfig{file=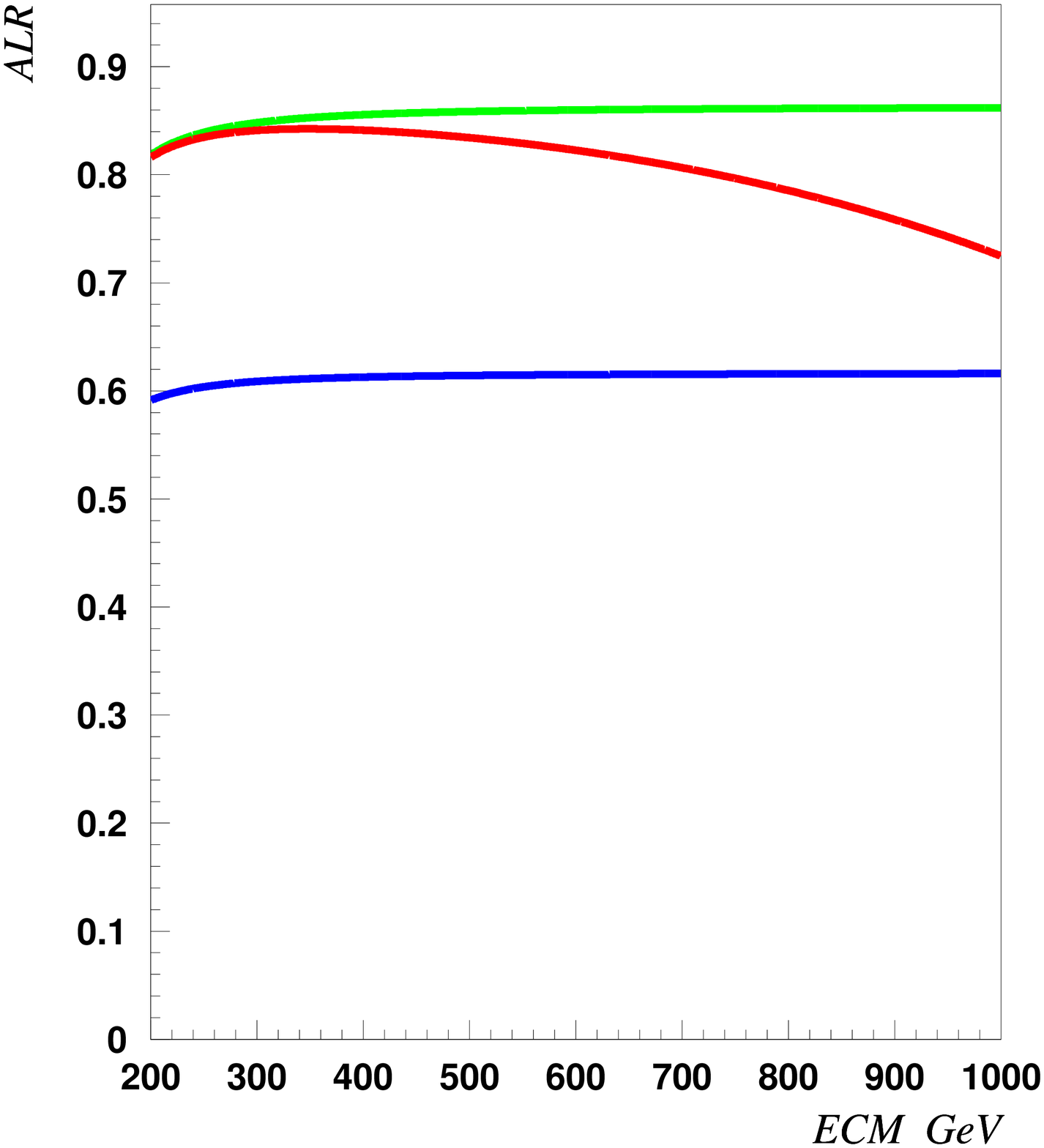,width=6.cm}
\epsfig{file=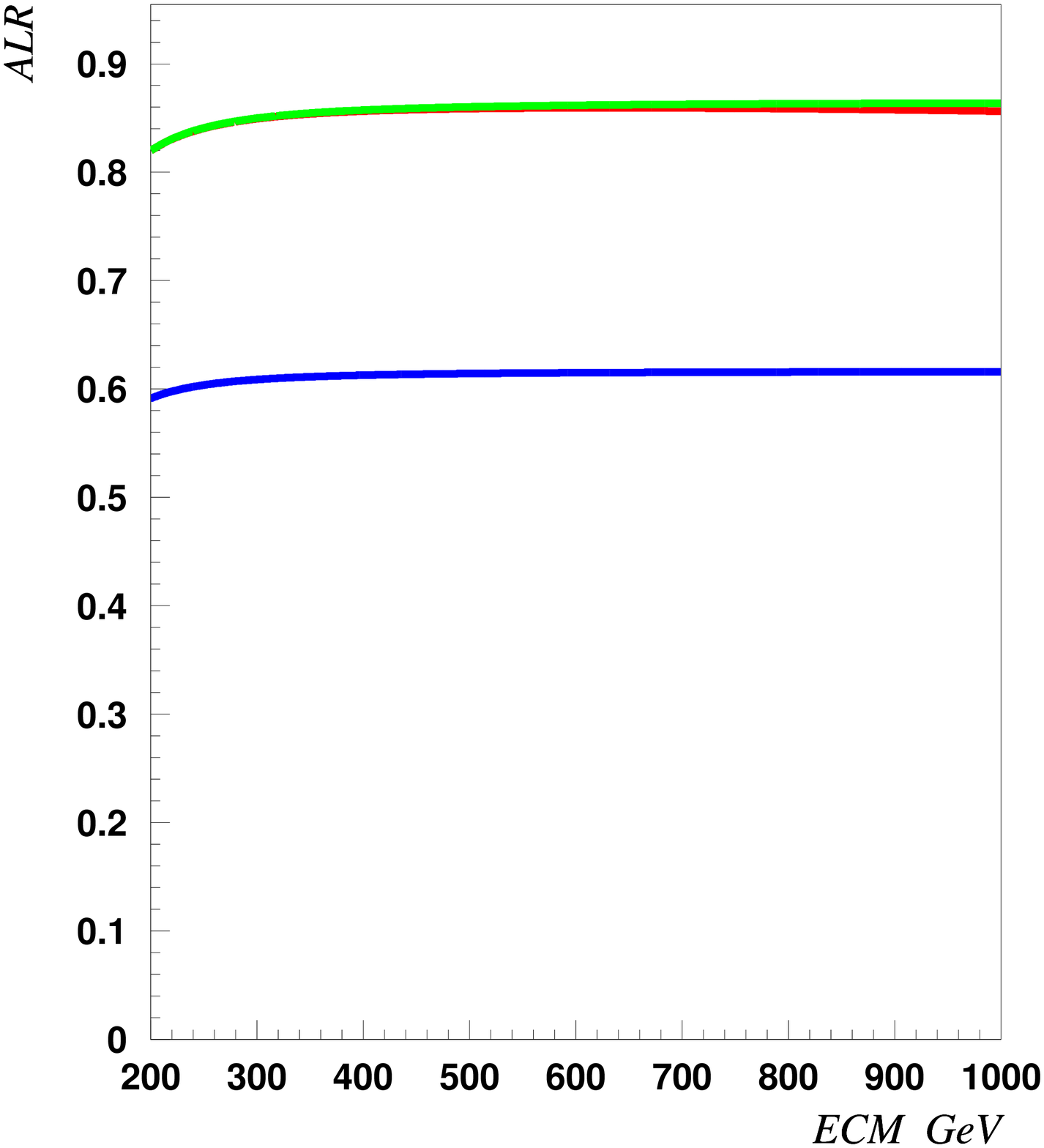,width=6.cm} 
\end{center}
\vspace*{-.5cm}
\caption{Energy dependence of the Left--Right polarization asymmetry for 
$b$--quarks in the SM and the RSa (left) and RSb (right) case including the  KK
contributions and in the pure mixing case. The blue curve is for the SM, the
green curve corresponds to mixing only while the red curve takes into account KK
exchange.}
\label{Abb-plot}
\vspace*{-.3cm}
\end{figure}

The lower part of Table 2 shows the accuracy of the $A_{LR}$ measurement with 
${\cal L}=500$ fb$^{-1}$; it is lower than at GigaZ but the effect is much 
larger as one can already notice at LEP2 energy from Fig.~\ref{Abb-plot}. The
study of the energy dependence of appropriate observables will allow to 
estimate the masses of the KK excitations. This is shown in Fig.~\ref{Abb-plot}
for the RSa case where one sees a clear difference between a pure mixing effect
[the flat  green curve] which dominates a low energies and the contribution of
the  KK states [the red curve which contains both contributions] which has an 
energy slope which allows to estimate the KK mass. If one uses the high energy 
data alone, without any theoretical input and without using the measurement of 
the mixing effect dominant at low energy, the error on the KK mass would be at 
the 10\% level for a 3 TeV KK state. One can perform much better by noting that
the  mixing effect is proportional to $F(c_{q_R})/(\sin\theta' M_{KK})^2$.  For
the heavy top quark, one can assume that $F(c_{t_R})$ is close to the 
asymptotic value, $\sim -8$, so that the mixing determines the denominator. 
This allows to extract the $c_{f_R}$ values for the electron and the $b$ quarks 
and thus the $Z'$ couplings; in this case one can estimate $M_{KK}$ to about  a 
percent. This argument indicates how ILC could allow to extract precisely  the
various parameters of the RS model.

\subsubsection*{5.2 Expectations for top quarks}

Excellent statistical accuracies are expected for the channel $\ee \to t\bar t$
since $\sim 4\times 10^5$ events can be collected with ${\cal L}=500$ fb$^{-1}$
at $\sqrt s=500$ GeV. A key issue will be the signal/background ratio which 
drives systematic errors of the measurements. Since top quark pairs are mostly 
made of six--jet events or four--jet events with one isolated lepton, with two 
of these jets being $b$ quarks, they can be very well separated from 
background events by using the following observations:
$i)$ $W^+W^-$ events which do not contain $b$--jets, have 4 jets and are very 
forward peaked;   
$ii)$ $ZZ$ events are an order of magnitude less numerous, have 4 jets and 
are forward peaked; 
$iii)$ $b\bar{b}$ pairs with multi gluons give an incompressible but small 
background;   
$iv)$ $ZW^+W^-$ events are an order of magnitude below the signal at 500 GeV; 
the channel $Z \to b\bar b$ can fake the signal topology but is at the \% level
only.\s

One can thus clearly tag the latter background and eliminate it further by 
reconstructing the $Z$ boson mass using the resolution anticipated with 
particle flow methods developed for ILC detectors which should lead to a few 
percent resolution on the $Z$ boson mass. This method will also allow to 
reconstruct the $W$ boson mass which will further reduce the 4 jet QCD 
background. Note also that selecting isolated leptons would decrease the 
efficiency by a factor of two but allows for a $\sim 100 \%$ purity. Therefore,
one can conclude that the statistical accuracy achievable at the ILC is
unlikely to be jeopardized by an insufficient knowledge of the background
contamination or of the signal efficiency. This, however, needs confirmation 
with appropriate simulation and reconstruction tools.

\begin{table}[!h] 
\vspace*{-.2cm}
\begin{center} 
\renewcommand{\arraystretch}{1.2}
\begin{tabular}{|c||c|c|c|} \hline  
Energy  &  Observables  & Error \%  & Deviation \%  \\ \hline
500 GeV &  $R_{\rm t}$ & 0.2 & +24   \\
 &  $A_{\rm LR}$ & 0.1 & +200  \\ \hline  
\end{tabular} 
\caption{Deviations and accuracies expected at ILC in the top sector assuming  
$\sqrt s=500$ GeV and ${\cal L}=500$ fb$^{-1}$ and electron polarization.}
\end{center}
\vspace*{-.7cm}
\end{table}

\begin{figure}[!h]
\begin{center}
\vspace*{-.3cm}
\epsfig{file=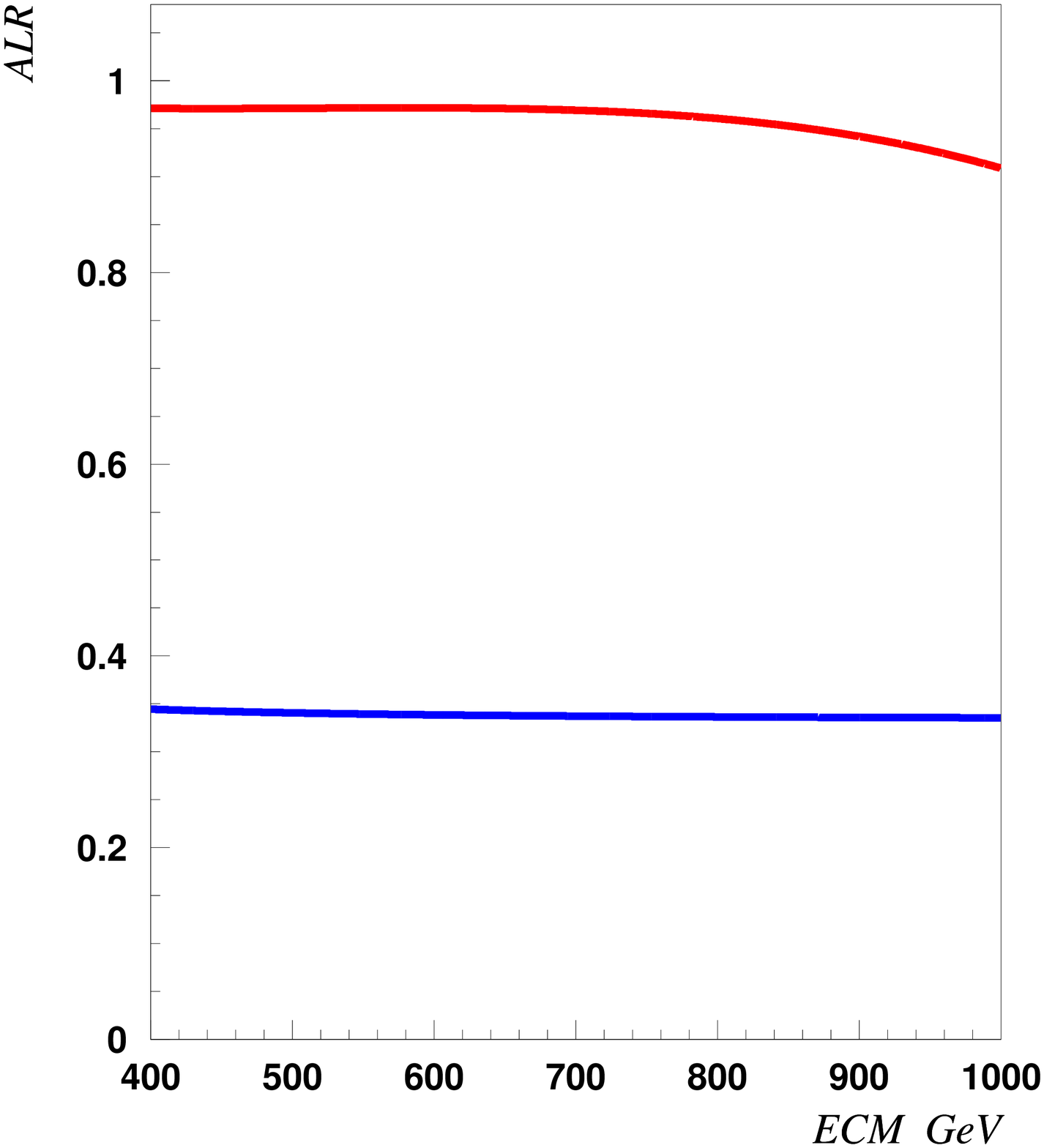,width=6.cm} 
\epsfig{file=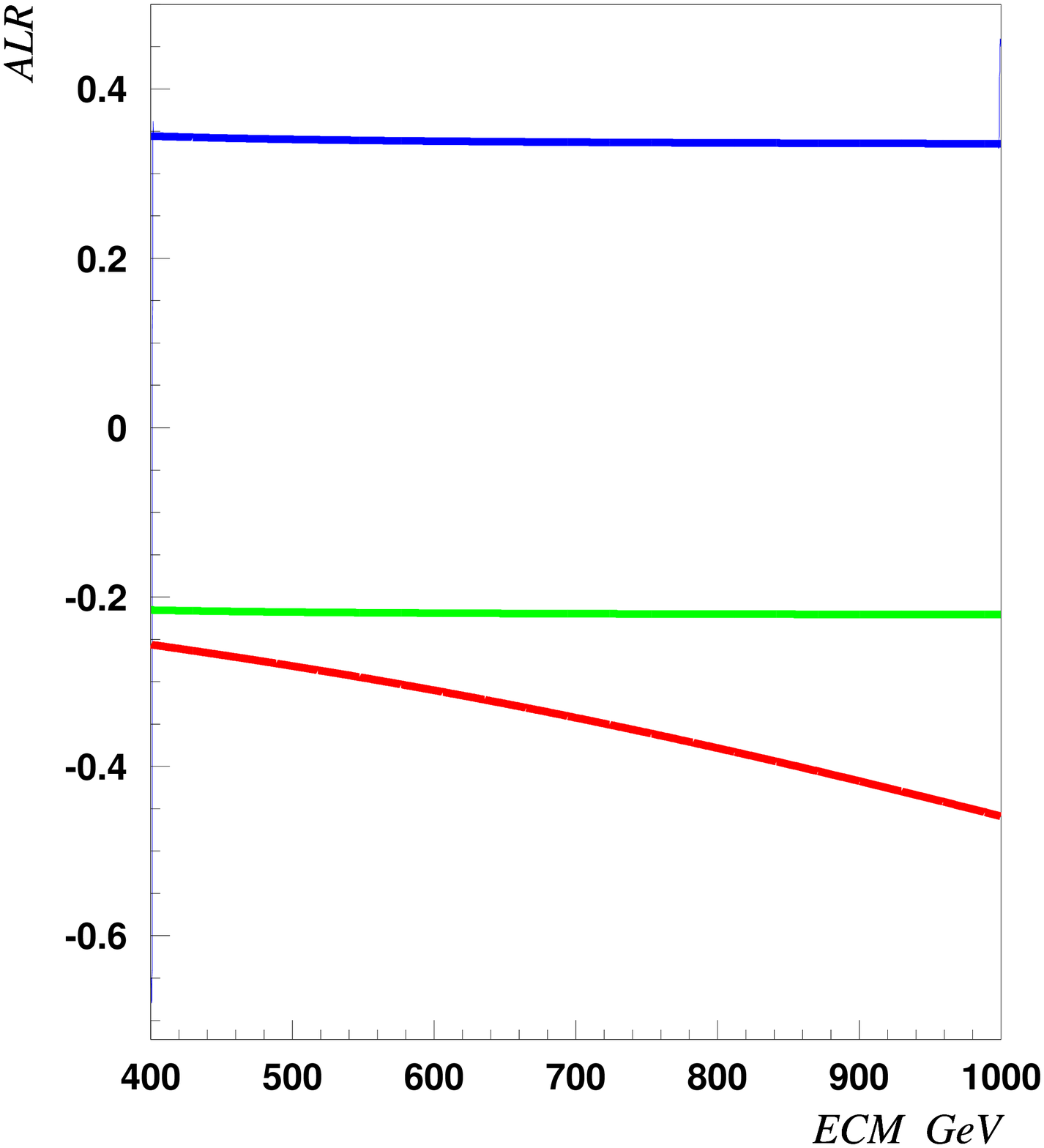,width=6.cm}
\end{center}
\vspace*{-.6cm}
\caption{Energy dependence of the Left--Right polarization asymmetry for top 
quarks in the SM and in the RSa and RSb scenarios. Conventions are as in the
previous figure.}
\label{Att-plot}
\vspace*{-.2cm}
\end{figure}

Table 3 displays the expected accuracies in the measurement of $R_t$ and 
$A_{LR}^t$ with $\sqrt s=500$ GeV and ${\cal L}=500$ fb$^{-1}$, with 
longitudinally polarized beams. Taking as an example the LR asymmetry, the
energy dependence of which is shown in Fig.~\ref{Att-plot}, a huge deviation  is
expected from the SM case, if one assumes $c(t_R) =-0.5$ as previously
discussed.  This means that given the accuracies expected at  the ILC, one could
as for GigaZ, allow for two orders magnitude reduction in  the KK effect and
still clearly observe a significant deviation. Thus, in the  top quark sector,
the sensitivity of the ILC is also sufficient to exclude  or confirm the RS
scenario; see also Ref.~\cite{Sher} for instance. \s 

\subsection*{5. Conclusion}

In this paper, we have analyzed the impact of a RS extra dimensional model with
the extended ${\rm SU(2)_L \times SU(2)_R \times U(1)_X}$ gauge structure
allowing relatively light KK excitations, on the high precision electroweak
data. In this model, in order to interpret the mass hierarchies,  the fermions
are localized differently along the extra dimension so that the electroweak
interactions for the heavy third generation fermions are naturally different
from those of light fermions.  The mixing between the $Z$ boson with its KK
excitations and those of the new $Z'$ boson, is then large enough to
significantly affect the $Z b\bar b$ vertex, while keeping the $Z$ coupling to
light fermions almost unchanged. We have shown that with a judicious choice of
the parameter assignments in the  $b$--quark sector and for order TeV KK states,
so that the gauge hierarchy problem is addressed,  the $\sim 3\sigma$
discrepancy of the forward--backward asymmetry of $b$--quark jets, $A_{FB}^b$,
with measurements performed on the $Z$--pole can be resolved, while all other
precision  electroweak data including the normalized $Z \to b\bar b$ partial
decay width  $R_b$, stay almost unaffected.\s 

The key point of the analysis is to generate a rather large contribution to the
right--handed $Zb_R\bar b_R$ charge which is small in the SM, to provide the
required negative shift in $A_{FB}^b$;  this contribution is annihilated by a
relatively smaller contribution to the left--handed $Zb_L\bar b_L$ charge,
leading to an almost constant $R_b$ value. In this context, we have analyzed two
possible scenarios. In a first scenario, in which the U(1) group is identified
with ${\rm U(1)_{B-L} }$, a  large contribution to the $Zb_R\bar b_R$ charge is
needed as to flip its sign.  However, at energies below and above the $Z$
resonance, although the experimental uncertainties are larger than at LEP1, the
theoretical predictions for $A_{FB}^b (s)$ are in a rather poor agreement with
the experimentally measured values. In a  second scenario, the charges of the
U(1) group have been chosen such that the  $b_R$ quark has isospin
$I_{3R}^{b_R}=+\frac12$ with respect to the new ${\rm SU(2)_R }$ group. In this
case, only a $\sim 30\%$ change of the $Zb_R\bar b_R$ charge  is needed to
resolve the $A_{FB}^b$ anomaly. In this scenario, the predictions for  the
asymmetry outside the $Z$--pole are also in a good agreement with the 
experimental data.\s

This scenario can be tested in detail at future high--energy colliders, and in
particular at the ILC. At GigaZ, i.e. when running ILC at energies close to the
Z--pole, the LEP1 and SLC precision measurements of $A_{FB}^b, A_e$ and $R_b$
can be vastly improved, as longitudinal polarization will be available and the
planed luminosities will be much higher. With the expected high accuracies, one
could allow for two orders magnitude reduction in the new effects and still
clearly observe a significant deviation. At high energies, $\sqrt s \gsim 500$
GeV, measurements in heavy quark production, in particular in the process $\ee
\to t \bar t$, will allow to perform precision measurement of the masses of the
KK states and their couplings to fermions. For instance, KK masses of the order
of 15--30 TeV can be probed and a mass $M_{KK} \sim 3$ TeV can be measured at
the percent level. A detailed discussion of the impact of this model at the LHC
and ILC will be given in a forthcoming publication \cite{DMR2}.

\vspace*{.5cm}

\noindent \textbf{Acknowledgments:} Discussions with Christophe Grojean and
Ritesh Singh are greatfully acknowledged; G.M. thanks K.~Agashe for useful
conversations.   The work of A.D. and G.M. is supported by the  ANR for
the project {\tt PHYS@COL\&COS} under the number NT05-1\_43598.


\end{document}